\documentstyle[12pt,twoside,fleqn,espcrc1,epsf]{article}

\newcommand{\AmS}{{\protect\the\textfont2
  A\kern-.1667em\lower.5ex\hbox{M}\kern-.125emS}}

\title{Strangeness Production and Ultrarelativistic Cascades}  

\author{D.~E.~Kahana\address{Physics Department, State University of New York at Stony Brook,\\
	Stony Brook, NY 11791, USA}%
	and
	S.~H.~Kahana
	\address{Physics Department, Brookhaven National Laboratory\\
   	Upton, NY 11973, USA}}

\begin{document}
\pagestyle{empty}
\maketitle  
  
\begin{abstract}
A two phase cascade, LUCIFER II\cite{LUCIFERII}, developed for the treatment
of ultra high energy ion-ion collisions is applied to the production of
strangeness at SPS energies $\sqrt{s}=17-20$. This simulation is able to
simultaneously describe both hard processes such as Drell-Yan and slower,
soft processes such as the production of light mesons, including strange
mesons, by separating the dynamics into two steps, a fast cascade involving
only nucleons in the original colliding relativistic ions followed,
after an appropriate delay, by multiscattering of the resulting
excited baryons and mesons produced virtually in the first
step. No energy loss can take place in the short time interval over which the
first cascade takes place.  The chief result is a reconciliation of the
important Drell-Yan measurements with the apparent success of standard
cascades to describe the nucleon stopping and meson production in heavy ion
experiments at the CERN SPS. A byproduct, obtained here in preliminary
calculations, is a description of strangeness production in the collision of
massive ions.
\end{abstract}   

\section{INTRODUCTION}  
Many cascades \cite{frithjof,werner,geiger2,RQMD,URQMD,ARC1,LUCIFERI}
have been constructed to consider relativistic heavy ion
collisions. As the eventual aim of experiments designed to study such
collisions is the creation of a regime in which the quark-gluon structure
of hadronic matter becomes evident, it is ultimately necessary to include the
partonic degrees of freedom in such cascades. However, since at SPS and even
at RHIC energies it is by no means clear that all initial or subsequent
hadron-hadron collisions occur with sufficient transverse momentum to free
all partons, at least a part of the eventual simulation must
deal with collisions of both the initially present baryons (nucleons in fact)
and in the end of all the produced hadrons (mesons) as well. In the course of
laying out the algorithms for the hadronic sector of the cascade a very
natural time-driven division between the hadronic and partonic sectors
arises.

\begin{figure}
\begin{minipage}[t]{80mm}
\epsfxsize=2.2truein\epsffile[29 70 580 800]{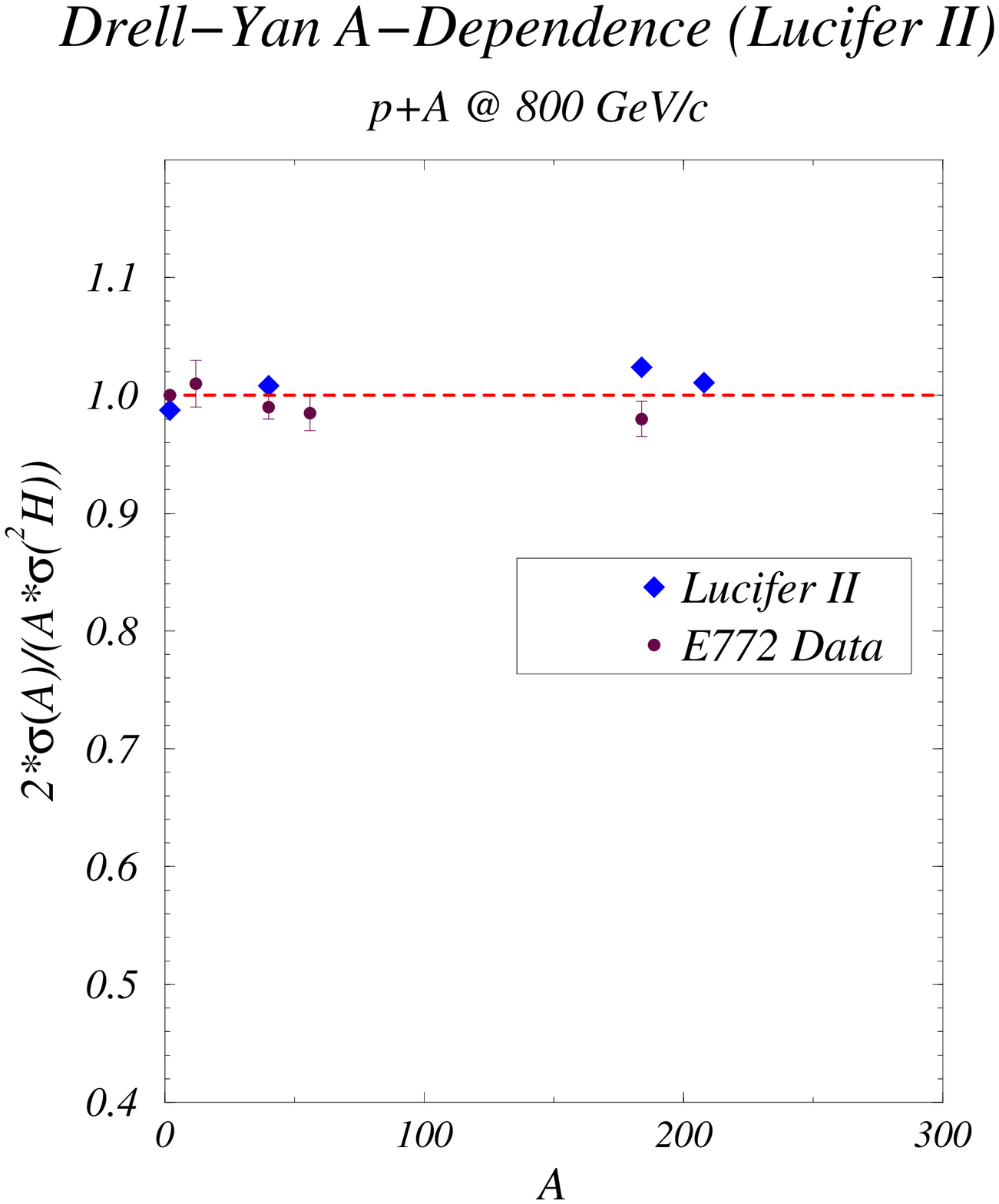}
\end{minipage}
\hspace{\fill}
\begin{minipage}[t]{80mm}
\epsfxsize=2.2truein\epsffile[40 90 570 800]{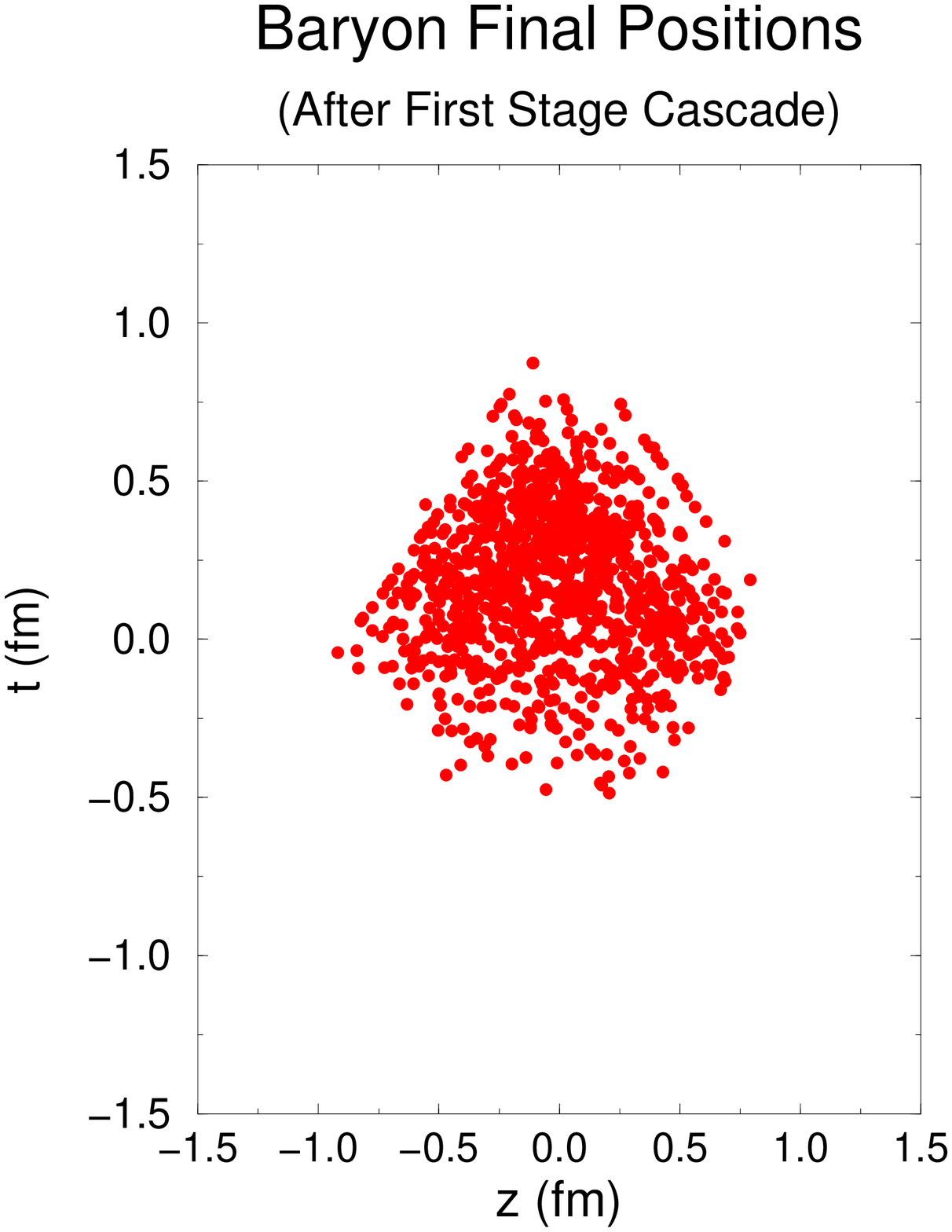}	
\end{minipage}
\caption[]{\label{fig:one}
(a) The A-dependence of measured Drell-Yan from E772 (FNAL) vs a Lucifer II
calculation, and (b) the nucleon-nucleon positions z vs t at the end of the 
first high energy cascade.}
\end{figure}

To begin with, however, we wish to discuss the global time scales which
divide the cascade into what one could designate as `hard'', perturbative 
partonic or `soft', non-perturbative or hadronic, processes.  This
separation has been often discussed in the literature
\cite{gottfried,koplik,amueller1,dokshitzer}, but
for our purposes here a general outline will suffice. Experimental results
\cite{earlypAdata} on both energy loss and forward pion production in high
energy proton-nucleus scattering were the key motivation for this reasoning.
Energy-loss and meson production at low transverse momentum $p_t$ will in
general represent `slow' processes.  To be contrasted are the `fast' or hard
processes of which Drell-Yan (DY) \cite{NA3,E772}, which we will
consider in here, is a good example. The time scales for these processes can
in the first approximation be inferred from the uncertainty principle. Small
momentum transfer collisions take long times $\sim p_t^{-1}$, while
production of lepton pairs with masses in excess of say $4$ GeV take much
less time, $\sim 4^{-1}$ GeV$^{-1}$. If one examines the Drell-Yan
data\cite{E772}, Figure \ref{fig:one}, the apparent $A$-dependence of the DY
cross-section ($\sigma \sim A$) in $p+A$ collisions, suggests that production
takes place only at the highest energy. The incoming proton simply counts all
the nucleons that it hits, and has an equal probability of producing a DY
pair in every collision. 

This stands in opposition to the expectation of a standard two body cascade,
in which successive collisions take place at lower and lower energies, and
which would predict that the cross-section rises more slowly than $A$.  These
results imply the incoming proton producing a DY pair, suffered no energy
loss in its passage through the nucleus. Nevertheless, calculations with a
purely hadronic cascade \cite{LUCIFERI} very well describe the energy loss
represented through the total pion spectrum seen in massive Pb+Pb collisions
at SPS energies. It is as if, though the protons in their initial
interactions inside a nucleus do not lose appreciable energy, they
nevertheless remember full well what number and sort of collisions they have
experienced. The question then is, can the initial energy retention but final
eventual loss be described in a unified dynamic fashion.

These apparently contradictory features can be put together in a single
particle or resonance-based multiscattering scheme, in a rather
robust and modular fashion\cite{LUCIFERII}.  The method simply consists of
running the cascade in two modes, a high energy fast-time stage in which the
collisions histories are recorded and only fast processes allowed to engage.
Using the entire space-time and energy-momentum history of this fast mode, a
reinitialisation of the cascade is performed and a second normal hadronic
cascading, at greatly reduced energy, carried out. Figure \ref{fig:one}(b)
shows the final positions of baryons in the first phase, and indicates the
almost light-cone like paths for the particles engaging in this initial
phase.

The intermediate reinitialization, inserted between the fast and slow
cacscades, is most critical and is generated from the detailed particle
collision and space-time history acquired during the initial high energy
cascading. The energy loss from the expected inelastic nucleon-nucleon
interactions for {\it any} nucleon can be computed, as well as the
entire trajectory of each particle. The procedure followed then is to
set up groups of nucleons, connected by similar local collision histories and
then use these groups in a conserving fashion to construct
the energy-momentum to appear in produced generic meson resonances. The
latter will be the principal cascading objects in the second low-energy
phase, along with of course the still present baryons from the first phase.

The model selected for describing elementary hadron-hadron collisions in the
soft cascade incorporates generic mesons and baryons, which are the agents
for rescattering. Indeed we might think of the generic mesons as consisting
of an excited bag of a constituent $q\bar q$ pair and the generic baryons as
similar objects constructed from three quarks.  The known details of
proton-proton collisions, from the initial energy down, provide information
on the multiplicity distributions, production mechanisms and
multiplicity-rapidity-energy loss correlations essential to the reconstruction
of the final baryon four-momenta and hence of the energy and momentum to be
deposited into mesons. The methodology followed is proceeds as closely as
possible by reconstructing two-body scatterings experienced for each nucleon
traced. Thus the fluctuations inherent in NN, in energy loss, multiplicity,
character (flavour etc.) can be mirrored in the reinitialisation.

One might refer to this two stage cascade as a separation into short and long
distance behaviour, a separation created by a factorisation of the time
scales, or equivalently by the momentum transfer involved in each
process. This feature has of course be much discussed in the literature
\cite{gottfried,koplik,amueller1,dokshitzer}.  This
approach has a highly beneficial effect on what one might fear most
in cascades, a possible strong frame dependence. Indeed, in
preliminary calculations at RHIC energy, $\sqrt(s)=200$, for a Au+Au b=0
collision, i.~e.~in our context a worst case scenario, one finds the
%(dn/dy)-{\pi^-}$ differs by only 10-15\% for the extreme cases of a lab
global frame and the center of mass, very moderate at RHIC's high energies
and virtually non-existent at the SPS. The reason is simple and generic. 
The first, high energy cascade, leads (Figure \ref{fig:one}) to particles
travelling uniquely on light cone paths. Any frame dependence then enters
mainly through the second phase cascading, which involves greatly reduced
energies.

The second phase involves generic resonances, both baryonic and mesonic, of
delta and nucleon, $\rho$, $\pi$ and $K$ character, with masses between $m_N$
and $1.6$ GeV for baryons and $0.3$ to $1.0$ GeV for normal mesons,
appropriately higher for strange mesons. That the non-strange generic meson
resonance mass should be centered near $600-700$ MeV should be no surprise,
and is in fact consistent with our simple picture for these objects, a
constituent $q\bar q$ bound pair.  The produced mesons are of course not
allowed to interact until some formation time $\tau_f$ has passed, this is
then one {\it real} parameter in the model, determinable perhaps from pA, or
from light nucleus collisions (SS) \cite{LUCIFERI}.  Finally these resonances
will decay into observable mesons, hyperons and nucleons. In this early work
we limit ourselves to $\pi$'s, and $K$'s.

In the following sections we lay out a brief description of (1), the basic
input to the cascading, i.~e.~the model for elementary two hadron interaction
(2), the treatment of nucleus-nucleus collisions and finally (3), the results
for production of both normal hadrons, protons and pions and strange
particles ($K$'s and $\Lambda$'s) in massive ion-ion collisions at the SPS.
For details of the cascade architecture we refer the reader to the previously
cited work\cite{LUCIFERII}.

\section{HADRON-HADRON INTERACTION MODEL}

The objective of the cascade approach to nucleus-nucleus collision is to
proceed from a knowledge of elementary hadron-hadron collision to a
prediction of the far more complex many body event. This is, as will be
clear, not completely possible in the environment of relativistic
collisions. It is not possible to ignore the time structure of the elementary
collisions nor the nature of the objects produced in the initial
particle-particle collisions, and which then partake in ensuing
collisions. At AGS momenta \cite{ARC1}, $\sim 1.0-5 GeV/c$, the introduction
of the most evident, lowest lying resonances, the $\Delta$, the $N^*$ and the
$\rho$, was often sufficient for a reasonably accurate picture of the
dynamics. The essential time structure for a many body collision was then set
by the resonance lifetimes, all $\sim 1.5$ fm/c. At the considerably higher
SPS and RHIC energies the Lorentz contractions are much more severe and it is
necessary to be guided by both experiment, in particular from pA and the
lighter ion-ion case, and existing theoretical considerations on
reinteraction of hadrons inside nuclei\cite{gottfried,koplik}. At these
higher energies we must of course also allow for the possibility of partons
being freed from hadrons in collisions of high enough transverse momentum. It
is nevertheless interesting to restrict this first work, aside from
Drell-Yan, to the collisions involving hadrons.

\begin{figure}
\hbox to\hsize{\hss
	\epsfxsize=2.2truein\epsffile[39 80 570 800]{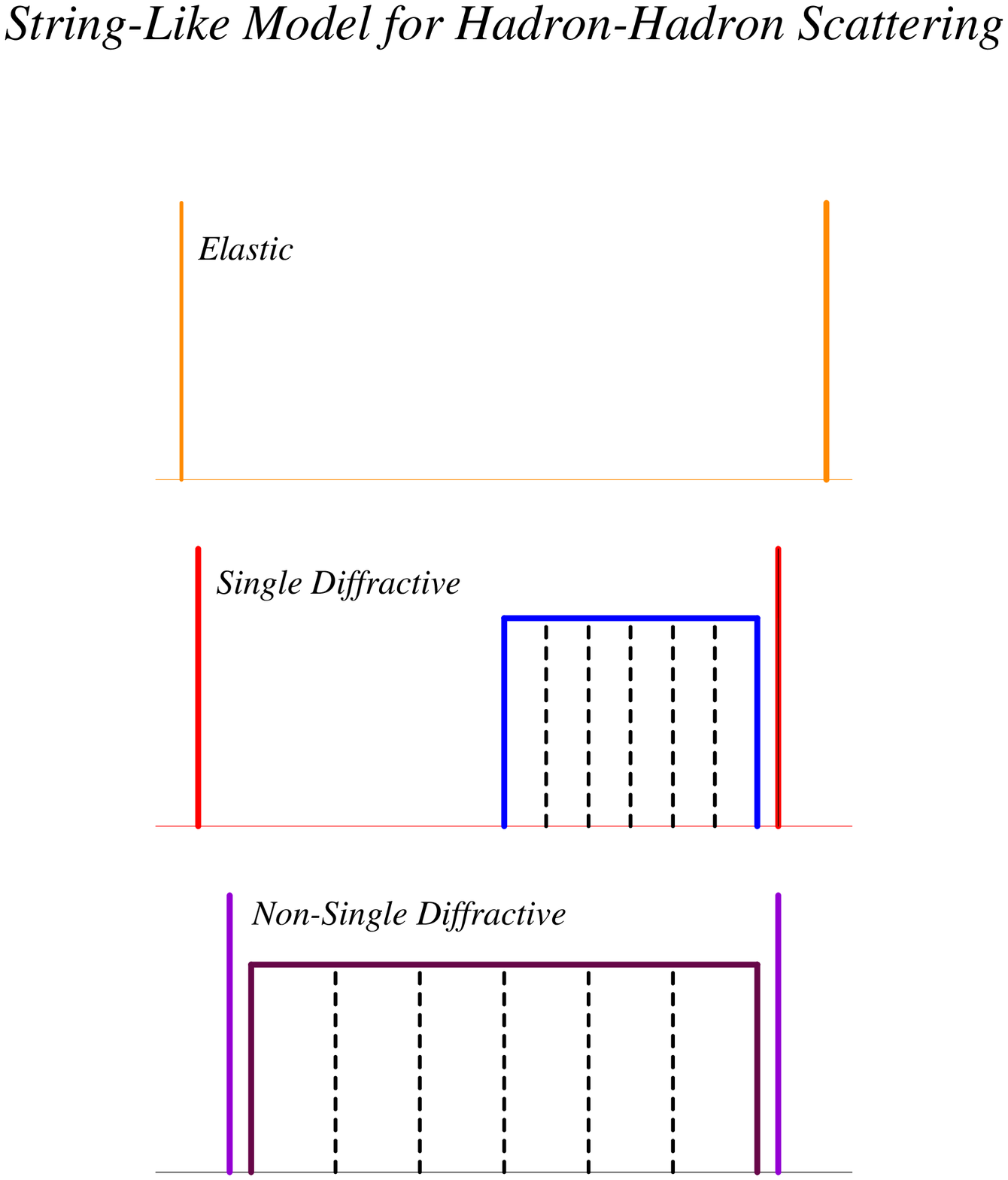}
	\hss
	\epsfxsize=2.2truein\epsffile[39 80 570 800]{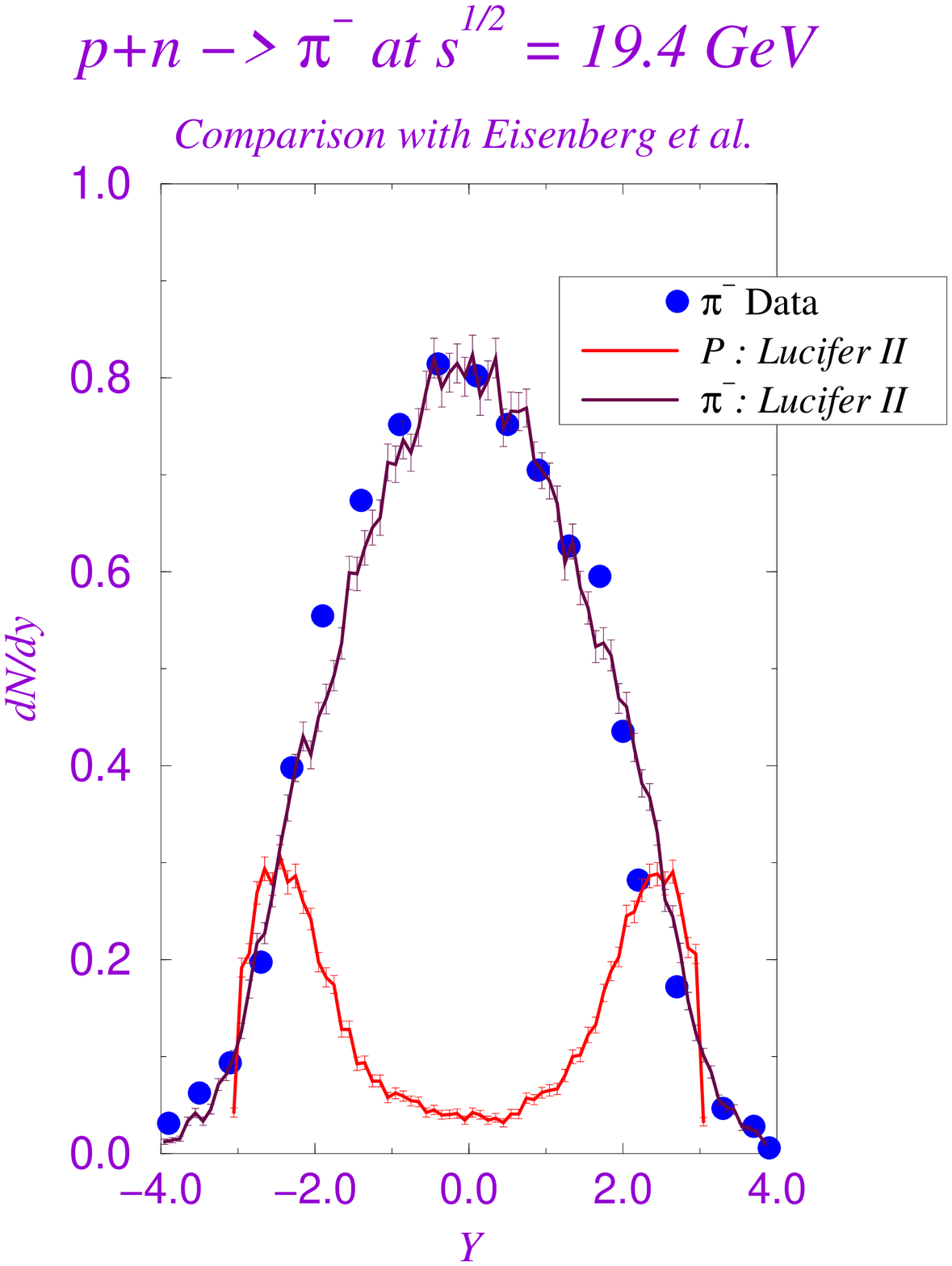}
	\hss}
\hbox to \hsize{\hss
  	\vbox{\hsize=2.2truein
	\hbox to \hsize{\hfil (a)\hfil}
	}\hss
  	\vbox{\hsize=2.2truein
	\hbox to \hsize{\hfil (b)\hfil }
	}\hss
	}
  \caption[]{\label{fig:two}
(a) Figurative representation of the elastic, single and non-single 
diffractive processes on which the elemntary hadron-hadron modelling is
based. The generic mesons are contained kinematically within the vertical
cross-hatched regions. Also(b) the fit obtained to known pn data near
the SPS energies.}
\end{figure}

\begin{figure}
\hbox to\hsize{\hss
	\epsfxsize=2.2truein\epsffile[39 80 570 800]{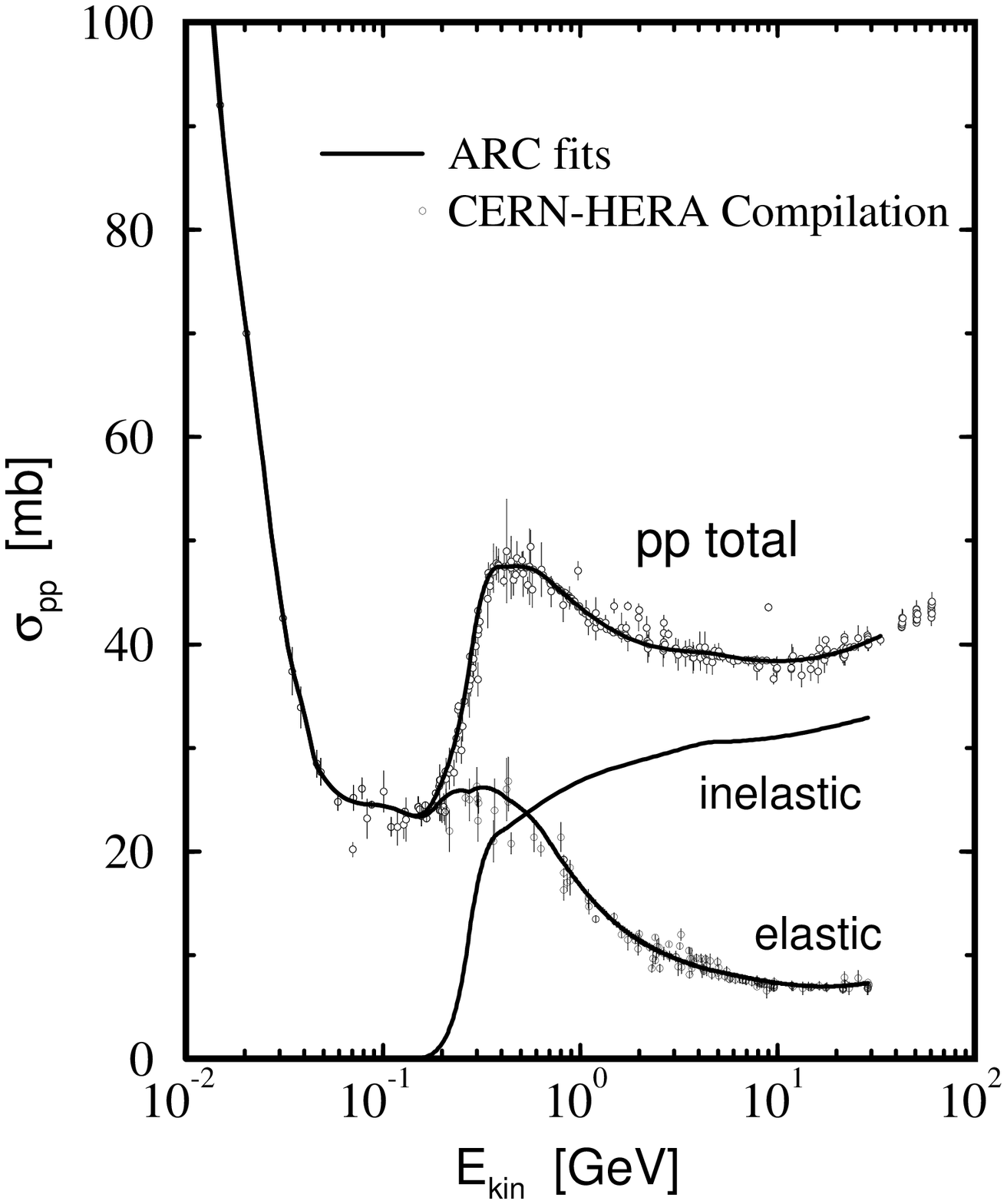}
	\hss
	\epsfxsize=2.2truein\epsffile[39 80 570 800]{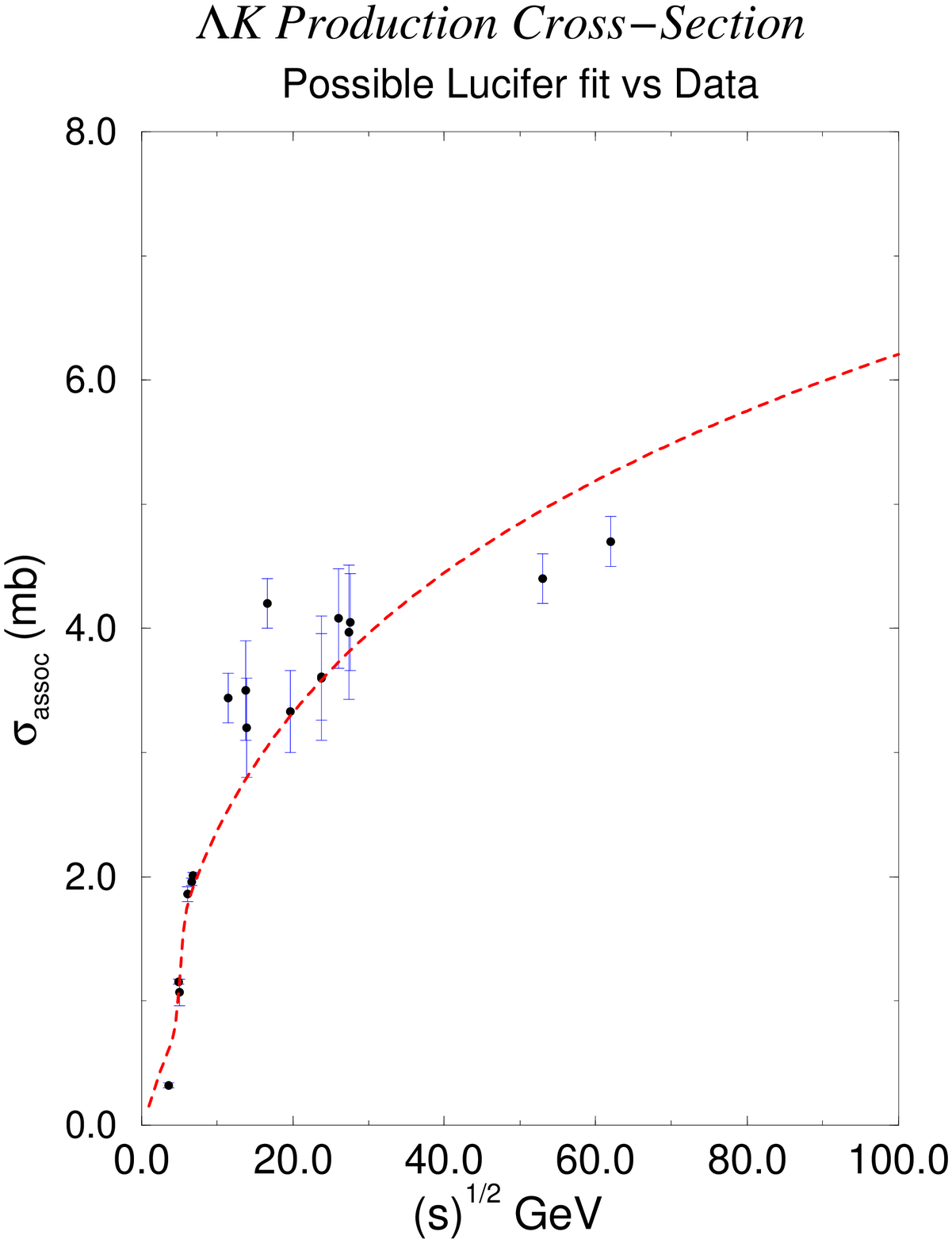}
	\hss}
\hbox to \hsize{\hss
  	\vbox{\hsize=2.2truein
	\hbox to \hsize{\hfil (a)\hfil}
	}\hss
  	\vbox{\hsize=2.2truein
	\hbox to \hsize{\hfil (b)\hfil }
	}\hss
	}
  \caption[]{\label{fig:three}
(a) The total proton-proton cross-section, and (b) the inclusive
	cross-section for associative $\Lambda+K$ production.}
\end{figure}

Many approaches have been put forward \cite{RQMD,URQMD}, some including
strings \cite{frithjof,werner}, but we try to retain a particle nature for the
cascade. The first element then is a model for the hadron-hadron system,
beginning with nucleon-nucleon but easily extended to meson-nucleon and
ultimately applied to any two body hadron-hadron collision.  The basic
processes are elastic scattering and inelastic production of mesons, the latter
divided into the well known categories\cite{goulianos} diffractive
scattering, referred to as single diffractive (SD), and non-single
diffractive (NSD)\cite{UA5}, figuratively displayed in Figure
\ref{fig:two}(a). The SD process, leading to a rapidity gap between one of
the leading hadrons and the produced mesons, is associated with a triple
Pomeron coupling \cite{pomeron}, while the NSD production is attributed to
single (and double) Pomeron exchange and presumably results in the observed
meson plateau. These diagrams then represent the basis for our development
but must be supplemented by an intermediate picture which allows us to apply
them, not only to hadron-hadron interactions in free space but also inside a
nuclear environment. The generic mesons depicted in Figure \ref{fig:two}(a)
and the generic baryons, with rather light masses selected in the ranges
suggested above, constitute the basic elements for rescattering in the
second-phase cascade. We reemphasize the $q\bar q$ and $qqq$ nature of the
generic resonances, structures easily related to the constituent quark models
frequently used for discussing soft QCD physics.  Two examples of the fits to
cross-sections we use are shown in Figure \ref{fig:three}(a) and in Figure
\ref{fig:three}(b) for the total pp and inclusive $\Lambda$-K respectively.

The soft Pomeron-mediated interactions \cite{pomeron} involve small $p_t$,
appreciable energy loss, and hence are thought to proceed on a very slow time
scale, $\sim 1$ fm/c in the rest frame of the relevant particles.  To the
generic mesons we also ascribe widths $\sim 125 MeV$ MeV.  All of the results
we obtain can in fact be well represented by using a single average mass for
the generic mesons near $600$ MeV. Strangeness bearing resonances, of course
produced associatively, are defined with masses and widths not unlike those
quoted above but appropriately higher. None of this two body information is
afterwards, in nucleus-nucleus interaction, adjusted, at least not so as to
destroy the fundamental connection with the two body measurements.

Care must be taken to describe the observed multiplicity distributions. We
have chosen to impose KNO scaling\cite{KNO} in our parametrisation of these
distributions. The results obtained for the two hadron system are very
similar to those in multiperipheral models\cite{koplik,amueller1,pomeron}.In
Figure \ref{fig:two}(b) we display the final fits to $pn$ at the highest
energies most relevant to this work \cite{Eisenberg}, $\sqrt(s)\sim 20 GeV$.
The total cross-section fit used was shown in Figure \ref{fig:two}(a).  The
generic resonances decay into two or three `stable' mesons, which include at
the moment, $\pi$, $\rho$ and $K$. There is of course a strong correlation
between the number of generic resonances formed and their mass, the latter
in turn determining the number of stable mesons produced by each generic
meson.

The two most important features of the input hadron-hadron system for the
ion-ion cascade are now set. These are (1) the energy loss and (2) the
multiplicities, in soft processes. Also included of course are the necessary
fluctuations in both these, from two body collisions and geometry, seen in
measurement. In particular, the degree of energy loss in SD and NSD, is
highly influential in determining the width in rapidity of the final
$\pi^-$ distribution in this figure. This in turn plays a direct role in
later allowing a sufficient and surprisingly high energy loss for protons in
the rather light collision of SS depicted in Section IV.  Many of the
nucleons in a central collision end up with low rapidity in the center of
mass\cite{NA35,NA49}, i.e. at mid rapidity.

\section{NUCLEUS-NUCLEUS COLLISION SIMULATION}

One of the evident puzzles we faced, and of course well known from
examination of both Drell-Yan \cite{NA3,E772} and the nucleus-nucleus
proton and meson data at the SPS\cite{NA35,NA49,NA50}, was the apparent
contradiction between the production of the massive lepton pairs on nuclei
and the energy loss\cite{LUCIFERI} seen in nucleus-nucleus. The first of
these latter processes is assumed hard and describable by perturbative QCD
(PQCD), while the latter typified by the production of mesons is assuredly
slow. We now try to exploit the differences in time scales, itself a function
of collision energy, to create a global cascade incorporating both
phenomena. To do this we separate the simulation into two phases, connected
by an intermediate reinitialisation. The first cascade records the entire
nucleon-nucleon history of the nucleus-nucleus collision. but permits energy
loss only from sufficiently rapid processes.

The methodology used in the first phase is straightforward and indeed in
outline closely resembles an eikonal or Glauber protocol while retaining the
random, and therefore fluctuating, collision nature of a cascade. The correct
cross-sections, at the incoming energy and employing Monte-Carlo, are used to
trace out the collision history. This history is then used to fix the
trajectory and final space-time positions of the nucleons. Importantly, the
recorded history can also be used to {\it properly} reconstruct and evaluate
the energy loss suffered by each nucleon. At the completion of this phase we
have in hand the number of collisions suffered by each baryon.

In the reinitialisation one introduces collision-related groups. It is into
these groups that the energy loss, history-determined particle by particle,
is placed. The group structures are virtually dictated by consideration of pA
where, in light of the relativistic $\gamma$'s at the SPS and at RHIC, the
incoming proton rapidly collides with a series of target nucleons in its
path. The energy-momentum lost is transferred onto the generic mesons, whose
multiplicity is also fixed by the collision record.  One may note the
resemblance to the wounded nucleon model\cite{woundednucleon}. Each incoming
nucleon has essentially been marked with its multifaceted history. One might
refer to this as a `painted' or programmed nucleon model.

The group selection procedure in the initial step is mostly topological in
nature. One first chooses the nucleon undergoing a maximum number of
collisions and continues, in a first pass, by associating with it the
nucleons with which it collided. Since we are considering only highest energy
collisions, these colliders are all going (in an equal velocity frame) in a
direction `opposite' to the originally chosen particle. Clearly we have begun
with particles near the colliding centers of the projectile and target.  One
more pass is made to augment and more importantly, kinematically to
symmetrise the groups. The particles added in the first pass are again
ordered by collision number, and the maximal collider in this ordering has
its `opposite' collidees included. 

The energy-momentum loss for each nucleon is tied to its collision history,
by reconstructing for each of its collisions, and with the use of the basic
elementary NN model of Section II, the specific losses and
multiplicities. This is done within the rest frame of the group, where also
the transfer of this energy-momentum is deposited onto the appropriate number
of generic mesons. The final distribution of these mesons in a group is
mostly dictated by the constraints of the conservation laws. Particular
problems of course occur at the edges of phase space in rare events.

Strangeness production, through associated $\Lambda-K$ or $K-\bar K$ choices,
is also allowed and in fact important at the SPS.  The baryon-baryon results
are then boosted back to the group rest frame. Transverse
momentum is added to the baryons as a random walk, according to collision
number, with basic distributions taken from the two-body model.

The final step in the reinitialisation is to place the mesons and baryons in
position and time, and to restart the cascade. The four-momenta, in the
global frame, of all particles are known as are the final positions and times
of the baryons. It was thought best to distribute the mesons in a group
randomly along the space-time path followed by each baryon in the initial
cascade (see Figure \ref{fig:one}(b)).

\section{LOW ENERGY CASCADE AND RESULTS}

In this section we describe the last stage of the cascade in a mode, not
unlike, but differing somewhat from the low energy cascade ARC \cite{ARC1}.
Only major meson states and resonances likely to significantly effect the
dynamics are included at this stage, i.e. the $\pi$, $K$ and $\rho$. Almost
without exception the ensuing collisions occur at relatively low center of
mass energy, certainly less than $\sqrt(s)\sim 5-8$ GeV. The low energy
cascadebegins only after the passage of a meson formation time, a parameter
presumably somewhat tuneable about the standard value $\sim 1$ fm/c.

\begin{figure}
\hbox to\hsize{\hss
	\epsfxsize=2.2truein\epsffile[42 115 620 720]{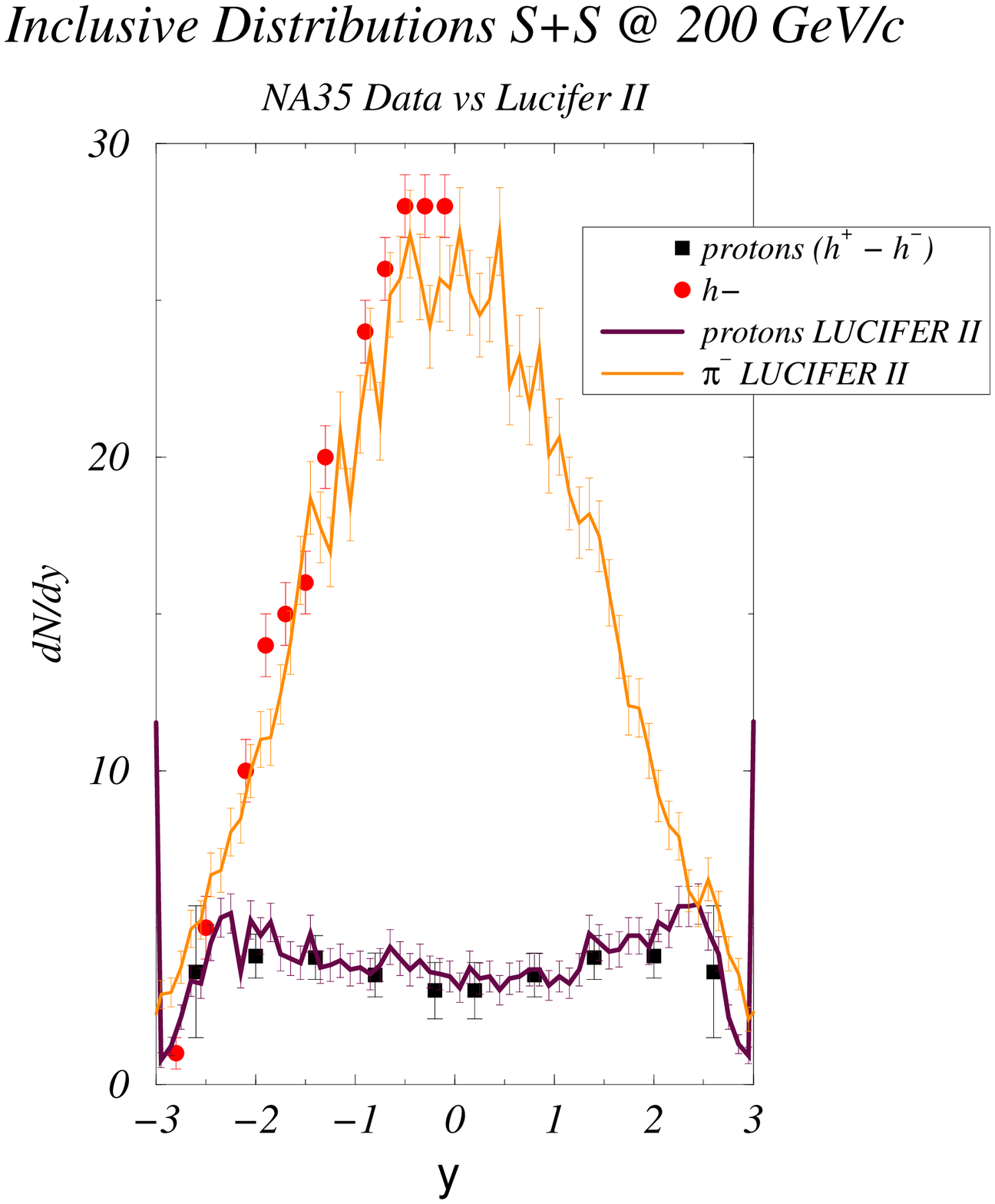}
	\hss
	\epsfxsize=2.2truein\epsffile[42 115 620 720]{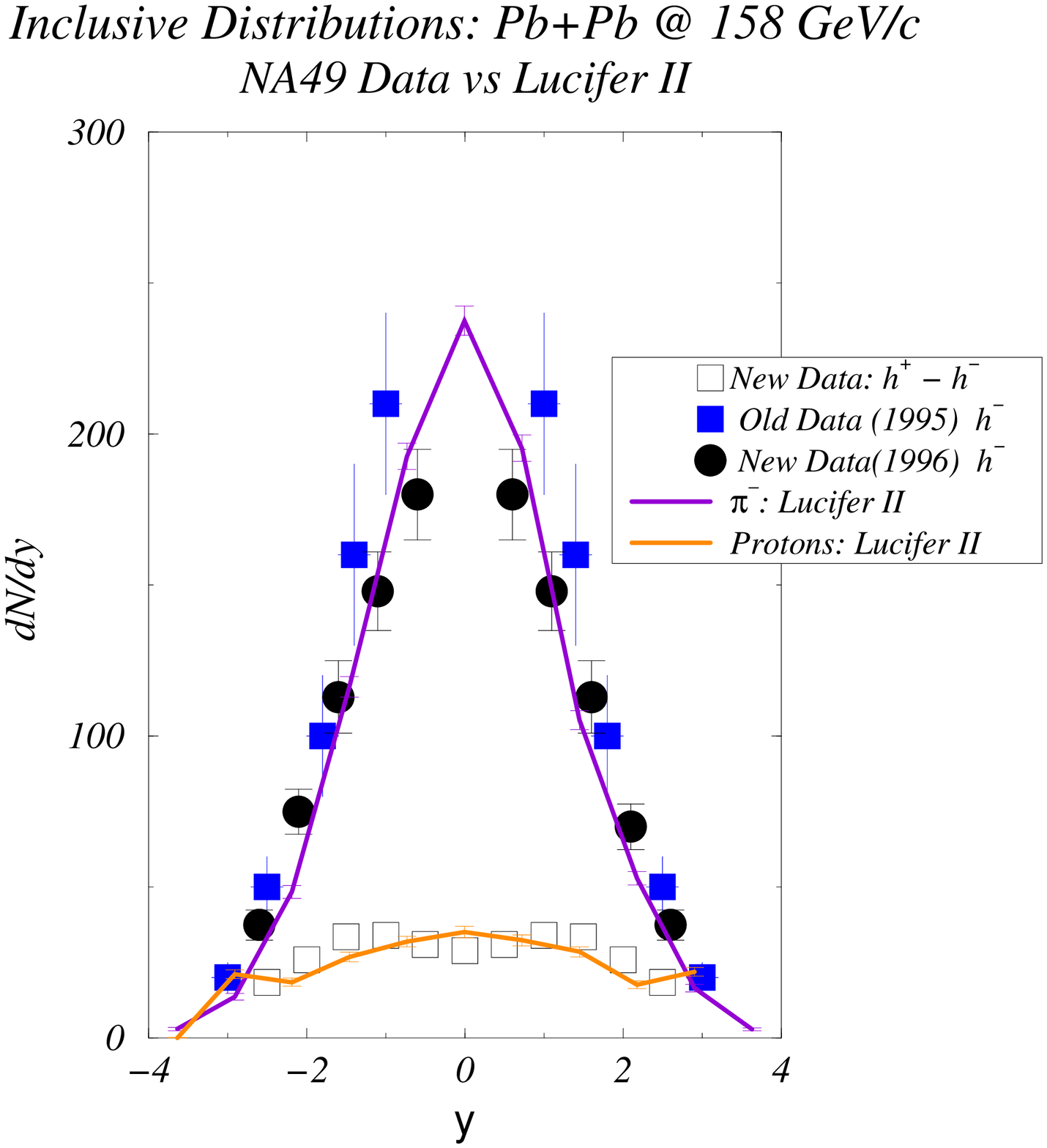}
	\hss}
\hbox to \hsize{\hss
  	\vbox{\hsize=2.2truein
	\hbox to \hsize{\hfil (a)\hfil}
	}\hss
  	\vbox{\hsize=2.2truein
	\hbox to \hsize{\hfil (b)\hfil }
	}\hss
	}
  \caption[]{\label{fig:four}
Rapidity distributions for $\pi^-$ and protons (a) compared to NA35 S+S data
and (b) compared to preliminary NA49 Pb+Pb data. In both experiments the
measurement of $\pi^-$'s is obtained from corrected $h^-$ and for protons
from $h^+$ - $h^-$.}
\end{figure}

The results of the low energy cascade, for a range of CERN SPS energies, are
displayed in Figures 4 and 5 for the SS and Pb+Pb systems. These represent
the output both high and low energy stages.  The calculated proton
distributions for S+S in Figure \ref{fig:four}(a) and Pb+Pb in Figure
\ref{fig:four}(b) exhibit the large, perhaps close to saturation energy loss
\cite{NA35,NA49}, suggested by the experimentalists already for the light SS
system. The corresponding pion spectra in SS and Pb+Pb are well reproduced,
in both magnitude and shape, by the Lucifer II simulation.  The average
number of collisions in central S+S is near 3.5 and closer to 7.5 in Pb+Pb,
with wide fluctuations about these averages seen for both nuclear
systems. Despite this difference in average collision number one can begin to
understand a saturation in stopping. At high energy considerable energy is
lost, perhaps one-half, per elementary collision and little is left after a
few collisions. However, the final soft cascading does rearrange and broaden
the rapidity distributions, though not much is added to the production of any
species.

\begin{figure}
\hbox to\hsize{\hss
	\epsfxsize=2.2truein\epsffile[39 80 570 800]{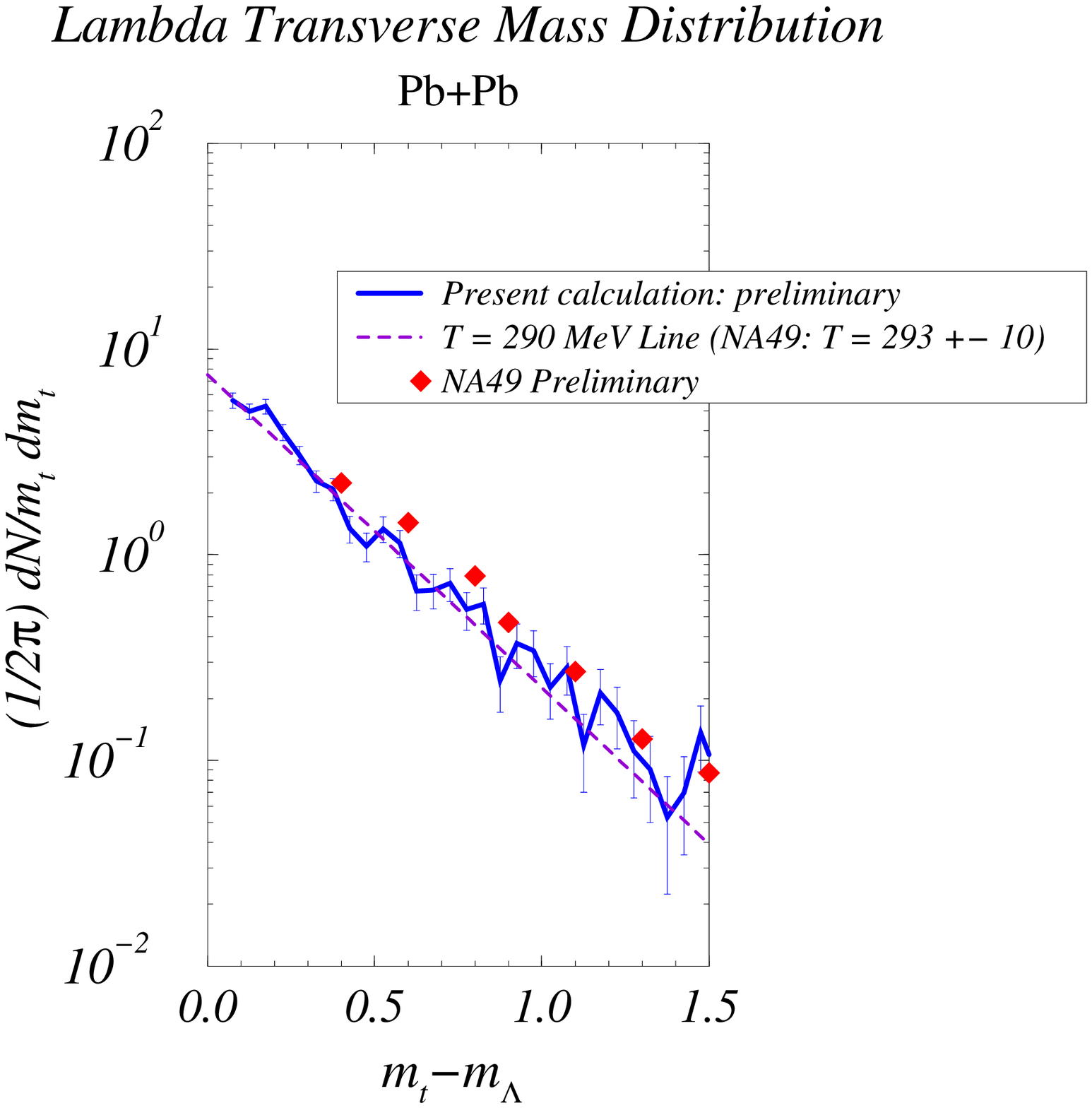}
	\hss
	\epsfxsize=2.2truein\epsffile[39 80 570 800]{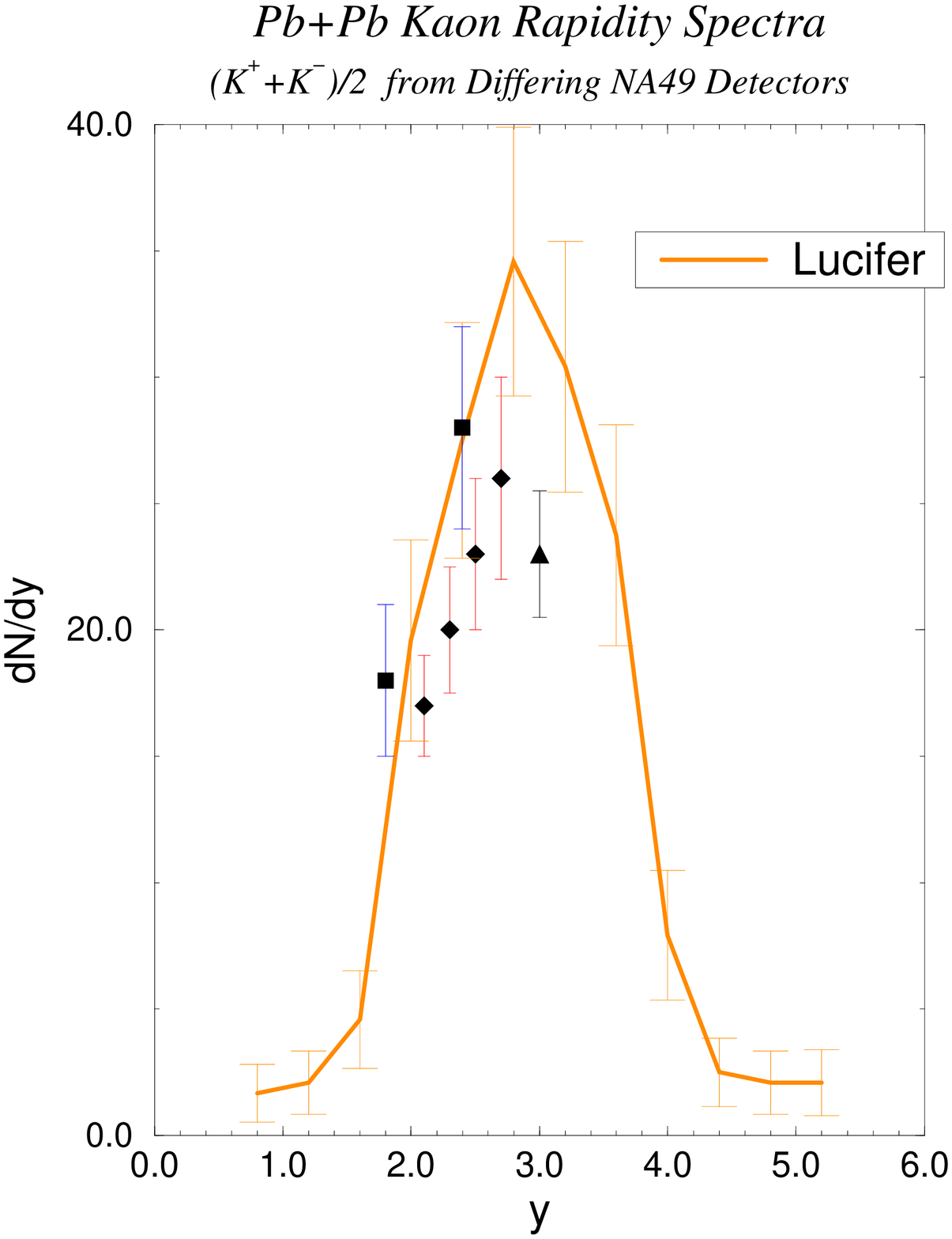}
	\hss}
\hbox to \hsize{\hss
  	\vbox{\hsize=2.2truein
	\hbox to \hsize{\hfil (a)\hfil}
	}\hss
  	\vbox{\hsize=2.2truein
	\hbox to \hsize{\hfil (b)\hfil }
	}\hss
	}
  \caption[]{\label{fig:five}
(a) Comparison of calculated transverse momentum distribution for $\Lambda$
	and also (b) of $K^+ + K^-$ production, to preliminary NA49 Pb+Pb
	data at SPS energies.}
\end{figure}

Strangeness production is here confined to results for Pb+Pb collisions at
$158$ GeV/c per nucleon incoming laboratory momentum. Theoretical
calculations for a transverse momentum $\Lambda$ spectrum and a $K$ rapidity
spectrum are compared to measurements in Figure \ref{fig:five}(a) and Figure
\ref{fig:five}(b). Elementary $K^+$ + $K^-$ is not so well known at the SPS
energies $\sqrt{s} \sim 20$ GeV and provides a degree of uncertainty to the
Pb+Pb prediction. One cannot ignore strangeness, sufficiently copious in the
most massive ion collisions to srongly affect the number of non-strange
mesons produced. That we are close to the measured rapidity densities for
both $K$'s $\pi$'s and $\Lambda$'s is a good sign. A more complete survey of
strangeness producing collisions, both for pA and AB must still be done.

It has of course been the principal thrust of this work to create a unified
dynamic approach, to ion-ion collisions at high energy, incorporating both
the results of Drell-Yan and the slower processes. So far, this has been done
allowing rescattering only via hadronic intermediate states. No partons have
been explicitly included, except insofar as structure functions must be used
for the calculation of Drell-Yan. This approach proves to be
phenomenologically successful, leading to a reasonable description of a broad
range of results obtained at the SPS. A secondary justification for following
such a pure hadronic calculation is to investigate the conclusions of
Kharzeev and Satz \cite{Kharzeev}, namely that a purely hadronic explanation
of the $J/\psi$ suppression seen by NA50 is not possible.  Certainly, the
energy and number densities for both baryons and mesons achieved in the
present simulations (LUCIFER II) are high enough $\sim 4-5$ GeV/$(fm)^3$ and
last for sufficiently long times for Pb+Pb at 158 GeV/c, to perhaps expect
unusual high density behaviour. The striking new results of {NA50}
\cite{NA50} must then be taken very seriously.

\section{ACKNOWLEDGEMENTS}
The authors are of course grateful to all of the experimentalists at
the AGS, SPS, ISR and FNAL, who have over the years gathered the basic
data. Certainly the code could not have been constructed without it. We also
wish to thank Y.~Dokshitzer and A.~H.~Mueller for instructive
conversations and advice. This manuscript has been authored under US DOE
contracts No. DE-FG02-93ER407688 and DE-AC02-76CH00016.


\begin{thebibliography}{9}
\bibitem{LUCIFERII}
Ultrarelativistic Cascades at SPS and RHIC energies I: Hadronic Sector, BNL-64948, (1997).
\bibitem{frithjof}
B.~Andersson, G.~Gustafson, G.~Ingleman, and T.~Sjostrand, Phys.~Rep.~97, (1983) 31.

\bibitem{werner}
J.~Ranft and S.~Ritter, Z.~Phys.~C27, (1985) 413;
K.~Werner, Z.~Phys.~C 42, (1989) 85.

%\bibitem{wang}
%X.~-N.~Wang and M.~Gyulassy, Phys.~Rev.~D44, (1991) 3501.

\bibitem{geiger2}
K.~Geiger and B.~Mueller, Nucl.~Phys.~B369, (1992) 600.

\bibitem{RQMD}
R.~Mattiello, A.~Jahns, H.~Sorge and W.~Greiner, Phys.~Rev.~Lett.~74, (1995) 2180.

\bibitem{URQMD}
H.~Stoecker, Proceedings, RHIC Summer Study~'96, D.~Kahana and Y.~Pang (eds.)
(1996).

\bibitem{ARC1}
S.~H.~Kahana, D.~E.~Kahana, Y.~Pang, and  T.~J.~Schlagel,
Ann.~Rev.~Nuc.~and Part.~Sc.~46, C.~Quigg (ed.~), (1996).

\bibitem{LUCIFERI}
D.~E.~Kahana, Proceedings, RHIC Summer Study~'96, D.~Kahana and Y.~Pang (eds.), (1996).

\bibitem{gottfried}
K.~Gottfried, Phys.~Rev.~Lett.~32, (1974) 957.

\bibitem{koplik} 
J.~Koplik and A.~H.~Mueller, Phys.~Rev.~D12, (1975) 3638.
                                                                      
\bibitem{amueller1}
A.~H.~Mueller, Proceedings, RHIC Summer Study~'96, D.~Kahana and Y.~Pang
(eds.), (1996).

\bibitem{dokshitzer}
Yu.~L.~Dokshitzer, Proceedings, RHIC Summer Study~'96, D.~Kahana and Y.~Pang
(eds.) (1996).

\bibitem{earlypAdata}
D.~S.~Barton et al., Phys.~Rev.~D27, (1983) 2580; 
W.~Busza et al., Phys.~Rev.~Lett.~34, (1975) 836.

\bibitem{NA3}
J.~Badier et al.~(NA3 Collaboration), Zeit.~Phys.~20, (1983) 101.

\bibitem{E772}
D.~M.~Alde et al.~(E772 Collaboration), Phys.~Rev.~Lett.~66, (1991) 133. 

%\bibitem{NA51}
%A.~Baldit et al.~(NA51 Collaboration), Phys.~Lett.~B332, (1994) 244. 


\bibitem{goulianos}
K.~Goulianos, Phys.~Rep.~101, (1983) 169.

\bibitem{UA5}  
G.~Ekspong for the UA5 Collaboration, Nucl.~Phys.~A461, (1987) 145c.

\bibitem{KNO}
Z.~Koba, H.~B.~Niesen and P.~Olesen, Nucl.~Phys.~B40, (1972) 317.

\bibitem{pomeron}
V.~N.~Gribov, B.~L.~Joffe and I.~Ya.~Pomeranchuk, Sov.~J.~Nucl.~Phys.~2, (1966) 549. 

\bibitem{Eisenberg}
Y.~Eisenberg et al., Nucl.~Phys.~B154, (1979) 239.

\bibitem{harris}
J.~W.~Harris, in Proceedings of the 12th Winter Workshop on Nuclear Dynamics,
Snowbird, Utah (1995).

\bibitem{NA35}
J.~Baechler for the NA35 Collaboration, Phys.~Rev.~Lett.~A461, (1994) 72.

\bibitem{NA49}
T.~Wienold and the NA49 Collaboration in Proceedings of Quark Matter~'96,
Nucl.~Phys.~A610, (1996) 76c; P.~G.~Jones and the NA49 Collaboration, ibid.
 
\bibitem{NA50}
M.~Gonin for the NA50 Collaboration, Proceedings of Quark Matter~'96,
Nucl.~Phys.~A610, (1996) 404c.


\bibitem{woundednucleon}
A.~Byalas et el., Nucl.~Phys.~B111, (1976) 461.

\bibitem{Kharzeev}
D.~Kharzeev and H.~Satz, In Proceedings of Quark Matter~'96,
Nucl.~Phys.~A610, (1996) 418c.
\end{thebibliography}
\end{document}